\documentclass%[aps,prl,superscriptaddress,nofootinbib,nobalancelastpage,showpacs,nobibnotes]{r%evtex4}
[aps,prl,twocolumn,superscriptaddress,nofootinbib,nobalancelastpage,showpacs,nobibnotes]{revtex4}
\usepackage{epsfig}

\begin{document}
\title
{Photoabsorption in a plasma in a high magnetic field   
}

\author{R. F. Sawyer}
\affiliation{Department of Physics, University of California at
Santa Barbara, Santa Barbara, California 93106}

\begin{abstract}
Photo-absorption in fully ionized plasmas in high magnetic fields 
is re-examined, using the methods of many-body quantum field theory.
For frequencies in the immediate vicinity of the electron cyclotron
resonance
the rates we obtain disagree markedly from those in the literature.
The new element in our work that causes most of the disagreement
is the inclusion of the lowest order real part of the energy-shift
of the resonant state, where, in effect, previous authors had included
only the imaginary part.

\end{abstract}
\maketitle

\author{R. F. Sawyer}
\affiliation{Department of Physics, University of California at
Santa Barbara, Santa Barbara, California 93106}

Photoabsorption in a plasma in the presence of a strong magnetic field has been addressed 
many times in the literature \cite{pp}-\cite{nv}. For the case of the dipole approximation to the absorption amplitude, and with the photon taken to be coupled only through the electron current, complete results can be found, for example,
in ref. \cite{pp}. Recently there has been attention given 
to the effects of the proton current in a hydrogen plasma \cite{pc}-\cite{zane2}. This becomes
important in regions of extreme field and low temperature
$\omega_{cp}/T \le 1 $, where $\omega_{cp}$ is the proton cyclotron frequency;
that is, when thermal photons have energy in or below the proton resonance
region. We will follow others' terminology and call this the region
of quantized protons.
The present paper has three main objectives:
\subparagraph{a.} To point out a complication that applies in either of the electron
or proton resonance regions. Qualitatively we describe it as follows:
In the resonance region photoabsorption is best described in relation to
resonant scattering. Then the usual free-free rate (times an easily calculated coefficient) becomes, in effect,
a contribution to the imaginary part of the self energy of the
propagator for the resonant state of the electron (or proton), to be added
to the imaginary part coming from the collisionless resonance
decay. Of course, there is a real part to the self energy part as well,
of the same order in the expansion parameter $e^2$ as the imaginary part.
Since it is small we might have assumed that the real part is just an 
inconsequential shift of the resonance energy. However, it is energy
dependent (and logarithmically singular at the position of the resonance).
We find that including the real part greatly changes both the shape of the spectrum near the
resonance and the integral over the resonance region.

\subparagraph{b.} To exhibit an approach to the ``quantized proton" problem
that we believe is superior to those found in the literature,
and to present a result for the ``Gaunt factor'' that is much simpler than
that contained in the appendex of ref. \cite{pc}.

\subparagraph{c.}
To address some collective effects that can be significant
in domains of higher density.
\newline

In discussing the issues that we shall raise, it is much more efficient, even for the recapture
of single electron results or classical results, to begin from a quantum field theory formulation. This formulation is at the same time
well adapted for addressing the emission, absorption and scattering processes that are involved,
and for incorporating the statistical mechanics of the plasma.
We consider non-relativistic spinless electrons, of mass $m$ annihilated by the field, $\psi_e ({\bf r})$,
and protons of mass $M$ annihilated by the field $\psi_p ({\bf r})$. We choose Coulomb gauge for
the electromagnetic field. Denoting the vector potential for the external field ${\bf A({\bf r})}$,
we define the current operator, as ${\bf j}({\bf r},t)=
{\bf j}_e({\bf r},t)+{\bf j}_p ({\bf r},t)$, where the currents
for the respective fields \{e,p\} are,

\begin{eqnarray}
 {\bf j}_e ={- e \over 2 m} \Bigr[ \psi_{e}^\dagger  [- i\vec \nabla - e \vec A ] \psi_{e} 
 +[ i\vec \nabla - e \vec A ] \psi_e^\dagger \,\,\psi_e \Bigr ]\,,
\nonumber\\ 
 {\bf j}_p ={ e \over 2 M} \Bigr[ \psi_{p}^\dagger  [- i\vec \nabla + e \vec A ] \psi_{p} 
 +[ i\vec \nabla + e \vec A ] \psi_{p}^\dagger \,\,\psi_{p}\Bigr ].
\label{currents}
\end{eqnarray}
We also define the number density operators $ n_{e,p}({\bf r})=\psi_{e,p}^\dagger ({\bf r})\psi_{e,p} ({\bf r})$.
We divide the Hamiltonian into an unperturbed part $H_0$, which includes all 
interactions with the externally applied magnetic field, 
\begin{eqnarray}
H_0= {1 \over 2m} \int d^3 r [i\vec \nabla +e \vec A ({\bf r})]\psi_e^\dagger  \cdot [- i\vec \nabla +e \vec A ({\bf r})] \psi_e 
\nonumber\\
+{1 \over 2M}  \int d^3 r [ i\vec \nabla -e \vec A ({\bf r})] \psi_p^\dagger  \cdot [- i\vec \nabla -e \vec A ({\bf r})] \psi_p  \,,
\end{eqnarray}
a Coulomb term, $H_c$, which
we divide into two pieces $H_c=H_c^{(A)}+H_c^{(B)}$,
\begin{eqnarray}
H^{(A)}_c = {e^2 \over 8\pi}\Bigr [ \int (d^3{\bf r}) (d^3{\bf r}') n_e({\bf r})
 {1 \over {|\bf r} - {\bf r}' |} n_e({\bf r}') 
\nonumber\\
+ \int (d^3{\bf r}) (d^3{\bf r}') n_p({\bf r})
{1 \over|{\bf r} - {\bf r}' |} n_p({\bf r}')\Bigr ]\,,
\end{eqnarray}
\begin{equation}
H^{(B)}_c =-{{e^2\over 4\pi}} \int (d^3{\bf r}) (d^3{\bf r}') n_e({\bf r})
{1 \over|{\bf r} - {\bf r}' |} n_p({\bf r}') \,,
\end{equation}
and a radiation term, $H_{\rm rad}$, which couples the external radiation to the matter,
\begin{eqnarray}
H_{\rm rad}=\int d^3 r\Bigr \{ \Bigr [{\bf j}_e+{\bf j}_p]\cdot {\bf A}_{\rm rad}
\nonumber\\
+e^2 \Bigr [{n_e \over 2 m} +
{n_p \over 2 M} \Bigr]{\bf A}_{\rm rad}\cdot {\bf A}_{\rm rad}\Bigr \}\,,
\end{eqnarray}

We choose the magnetic field ${\bf B}$ to be in the $\hat z$ direction.
In this case it is convenient to define currents
$j_{e,p}^{\pm}=([j_{e,p}]_x \pm i [j_{e,p}]_y)/\sqrt{2}$ which couple photon polarization vectors of
the form $(1 ,\mp i ,0)/\sqrt 2$.
In the dipole approximation to photon absorption and emission
 we shall encounter the space integrals of the currents (\ref{currents})
\begin{eqnarray}
{\bf J}_{e,p}=\int d^3 r \, {\bf j}_{e,p}({\bf r})\,.
\end{eqnarray}
In a system governed by $H_0$ each particle in the ensemble 
of electrons and protons
moves independently and the
second quantized formalism is unnecessary. For a single electron system, for example,
the action of the operator ${\bf J}_{e}$ on a state is the same
as the action of the operator $e \Pi=e({\bf p}-e{\bf A})$ on the wave function
for that state.
Thus the familiar relations for an electron in a constant magnetic field,
$[\Pi_e^+,\Pi_e^-]=eB=m \omega_{ce}$,  $[\Pi_e^\pm,H_0]=\pm \omega_{ce}\Pi_e^\pm$, and
$[\Pi_p^+,\Pi_p^-]=-\omega_{cp}$,  $[\Pi_p^\pm,H_0]=\mp \omega_{ce}\Pi_p^\pm$,
 translate into,
\begin{eqnarray}
[J^\pm_e,H_0]=\mp \omega_{ce} J_e^\pm~~,~~[J^\pm_p,H_0]=\pm \omega_{cp} J^\pm_p\,.
\label{basiccom}
\end{eqnarray}
where $\omega_{ce}$ and  $\omega_{cp}$ are the respective electron and proton
cyclotron frequencies. 
When we introduce the Heisenberg picture in the usual way,
with $H=H_0+H_c+H_{\rm rad}$, the equations (\ref{basiccom}) give,
\begin{eqnarray}
\Bigr ( {i\partial \over \partial t} \pm \omega_{ce}\Bigr )J_e^\pm (t)=
[J_e^\pm (t),(H_c(t)+H_{\rm rad}(t))]\, ,
\nonumber\\
\Bigr ( {i\partial \over \partial t} \mp \omega_{cp}\Bigr )J_p^\pm (t)=
[J_p^\pm (t),(H_c(t)+H_{\rm rad}(t))]\, .
\label{key}
\end{eqnarray}
These are the key equations for our application.
Turning to the commutators of the $J$'s with the Coulomb Hamiltonian, it
is more transparent if we use the 3D vector representation and express the ${\bf J}$'s in terms of 
infinitesimal translation operators for electron coordinates ${\bf T}_{e}$, and for
proton coordinates, ${\bf T}_{e}$,
\begin{eqnarray}
{\bf  J} _e={-ie \over m}{\bf T_e}+e^2\int d^3r\, n_e({\bf r}) {\bf A}({\bf r})\,,
\nonumber\\
{\bf  J} _p={ie \over M}{\bf T_p}+e^2\int d^3r \,n_p({\bf r}) {\bf A}({\bf r}),
\label{trans}
\end{eqnarray}
where,
\begin{eqnarray}
[{\bf T}_{e,p},\psi_{e,p} ({\bf r})]= {\bf \nabla}  \psi_{e,p} ({\bf r})\, .
\end{eqnarray}
Noting that  $[n_{e,p}({\bf r}) {\bf A}({\bf r}), H_c]=0$ we find first that,
\begin{equation}
[{\bf J}_{e,p}, H_c^{(A)}]=0 \, ,
\label{trans2}
\end{equation}
since the e-e and p-p interaction terms are separately invariant under the separate translations 
of either the electron or proton coordinate. However the e-p interaction $H_c^{(B)}$ is invariant
only under simultaneous translations for the electrons and protons,
\begin{eqnarray}
[{\bf T}_e+{\bf T}_p, H_c]=\Bigr [ \Bigr( {m \over ie} {\bf J}_e - {M \over ie} {\bf J_p}\Bigr ),H_c]=0 \, .
\label{trans3}
\end{eqnarray}
The separate commutators needed in (\ref{key})are thus,
\begin{eqnarray}
[{\bf J_e},H_c ] = -{M \over m}[{\bf J_p},H_c ]~~~~~~~~~~~~~~~~~~~
\nonumber\\
=i{ e^2 \over 4 \pi m}\int (d^3{\bf r}) (d^3{\bf r}')\, 
 \nabla_{\bf r} n_e({\bf r},t)
{1 \over |{\bf r} - {\bf r}' |} n_p({\bf r}',t) \,.
\label{ ccom}
\end{eqnarray}

\section{Formal expressions for the photo-absorption rate.}
Since the issues raised in the present paper pertain largely to the modes polarized
perpendicularly to the magnetic field, we address only the absorption of these modes.
We begin from the expression that the field theoretic formulation of statistical mechanics provides for the photo-absorption rate $\gamma_A(\omega)\pm$ in the medium, where $\pm$ refers
to the photon modes with polarization vector $(1,\pm i,0)\sqrt 2$

\begin{eqnarray}
&\gamma_A(\omega)^\pm= \frac{1}{2\omega  
[Vol.]}\int d^3 x  \, d^3y \,  dt \, e^{i\omega t} e^{-i \bf q \cdot(x-y)} 
\nonumber\\
&\times\langle    j^\pm ({\bf x}, t)  j^\mp \rm ({\bf y},0)
\rangle \,.
\label{photoabs}
\end{eqnarray}

The Heisenberg operators $j^{\pm}(x,t)$ are the relevant combinations of the total electromagnetic current operator,
$j^{\pm}=([j_e+j_p]_1 \pm i[j_e+j_p]_2)/ \sqrt 2$. 
The brackets, $\langle \rangle$ stand for thermal average, $\langle A \rangle= \rm {Tr}[e^{- \beta H}A]/{Tr}[e^{- \beta H}]$ where $\beta=T^{-1}$. 
Taking the dipole limit, $q \rightarrow 0$, and using the definition $J_{e,p}^\pm=\int d^3r j^\pm_{e,p}({\bf r})$
we obtain, after an integration by parts,
\begin{eqnarray}
&\gamma_A(\omega)^\pm = {1 \over 2\omega} \int dt \, \, e^{i\omega t} 
{1 \over \omega\pm \omega_{ce}}\Bigr ( {i\partial \over \partial t} \pm \omega_{ce}\Bigr )
\langle
 J_e^\pm (t) \,  j^\mp({\bf 0},0)  \rangle
\nonumber\\
&+  {1 \over 2\omega}\int dt \, \, e^{i\omega t} 
{1 \over \omega \mp\omega_{cp}}\Bigr ( {i\partial \over \partial t} \mp \omega_{cp}\Bigr )
\langle J_p^\pm (t) \, j^\mp({\bf 0},0)  \rangle \, .
 \label{dipole}
\end{eqnarray}
Now we substitute (\ref{key}), letting $A^{\rm rad}=0$ since (at the moment) we are 
considering a process in which no photons are involved except the one being absorbed.
\begin{eqnarray}
&\gamma_A(\omega)^\pm ={1\over 2 \omega} \int dt \, \, e^{i\omega t} 
\langle \Bigr [ \Bigr \{ {J_p^\pm (t) \over \omega \mp \omega_{cp}}+{J_e^\pm (t) \over \omega 
\pm \omega_{ce}}
\Bigr \},H_c(t)\Bigr ]
j^\mp(0)\rangle  
\nonumber\\
&={-1\over 2 \omega} \int dt \, \, e^{i\omega t} 
\langle   \Bigr [ \Bigr \{ {j_p^\mp ({\bf 0},0) \over \omega\mp \omega_{cp}}+{j_e^\mp ({\bf  0},0) \over \omega \pm \omega_{ce}}
\Bigr \},H_c (0)\Bigr ]J^\pm (t)\rangle \, .
\nonumber\\ \,
\end{eqnarray}
The second form follows from translational invariance (to shift the space integral from the first to the second current), space inversion (introducing a (-) sign)
and time translational invariance (a displacement $-t$) in the thermal average
factor, followed by complex conjugation.
Performing the same steps once again we obtain,
\begin{eqnarray}
&\gamma_A(\omega)^\pm 
={-1\over 2\omega} \int dt \, \, e^{i\omega t} 
\langle  \Bigr [\Bigr ( {J_p^\mp (t) \over \omega\mp \omega_{cp}}
+{J_e^\mp (t) \over \omega\pm \omega_{ce}}
\Bigr ) ,H_c(t)\Bigr ] 
\nonumber\\
&\times \Bigr [ \Bigr ( {j_p^\pm ({\bf 0},0) \over \omega\mp \omega_{cp}}+{j_e^\pm ({\bf 0},0)
 \over \omega\pm \omega_{ce}}
\Bigr ),H_c(0)\Bigr ]\rangle \, .
\label{both}
 \end{eqnarray}
Using (\ref{trans}) and (\ref{trans2}) we can rewrite (\ref{both} )in a simpler form\footnote{ The volume, Vol. , enters or leaves our formulae depending on whether
or not we use, e.g.,  the current evaluated at an arbitrary point in space, $j_e^\mp (0,0)$, as in
(\ref{both}) or the space integral of the current which, as it occurs in (\ref{2com}), is just the
current at the point times the volume, in view of the translational invariance of the thermal average
factor in (\ref{both}).} 
\begin{eqnarray}
\gamma_A(\omega)^\pm 
={1\over  2\omega [{\rm Vol.}]}\Bigr({1 \over M(\omega\pm \omega_{cp})}
+{1\over m(\omega \mp \omega_{ce})}\Bigr )^2
&\nonumber\\ \times
 \int dt \, \, e^{i\omega t} 
\langle  \Bigr [ T_e^\pm (t)
 ,H_c (t) \Bigr ] 
\Bigr [   T_e^\mp (0)
,H_c(0)\Bigr ]\rangle \, .
\nonumber\\ \,
\label{2com}
 \end{eqnarray}
or explicitly, using (\ref{trans3}),
\begin{eqnarray}
&\gamma_A(\omega)^\pm = {e^6 \over 32 \pi^2\omega  }
\Bigr[{\omega \over m(\omega\pm \omega_{cp}) (\omega \mp \omega_{ce})}\Bigr ]^2
\nonumber\\
&\times \int (d^3{\bf r}_1) (d^3{\bf r}_2)
(d^3{\bf r}_3) \,
 { 1 \over  |{\bf r}_1-{\bf r}_2| }  
\, { 1 \over |{\bf r}_3 | }
\int dt \, e^{i\omega t}
\nonumber\\
&\times \left\langle
\partial_\pm n_e({\bf r}_1,t)n_p({\bf r}_2,t)  
 \partial_\mp n_e({\bf r}_3,0) n_p(0,0)  
\right\rangle \,.
\nonumber\\ 
\label{dipoleee}
\end{eqnarray}
where we have defined,
$\partial_\pm \equiv (\partial/\partial x \pm i \partial/\partial y)/{\sqrt 2}$, and simplified the 
prefactor using $M\omega_{cp}=m\omega_{ce}$ and \newline $(m+M)/M\approx 1$.
This prefactor in (\ref{dipoleee}) agrees with that found in ref. \cite{pc}, except for the
damping terms that matter only very near the 
resonance in the results of these authors. This subject will be discussed at length in secs. 4 and 5. 

To recapitulate, (\ref{dipoleee}) gives the exact rates in the dipole limit. All of the effects
of Coulomb interactions in the medium are included. Of course, the four-density
correlator must calculated in an 
approximation. There are different domains in which it makes sense to take
different approaches. For example, at high temperatures and low densities we can use
the result in the absence of Coulomb coupling (beyond that already exhibited explicitly
in (\ref{dipoleee})). For low temperatures and low densities we
need to include full Coulomb wave-functions in the 
electron's interactions with a single ion. For high densities collective effects
become interesting, and  ``ring approximation" sums required (at the least).

\section{Born approximation in two regions}
The four-point density correlator depends on both Coulomb and magnetic
interactions. By ``Born approximation" we mean
that the Coulomb couplings are turned off in calculating the
bracket in (\ref{dipoleee}). In this case the correlator
factors into an electron part and a proton part, separately translationally invariant,
\begin{eqnarray}
\left\langle
\partial_\pm n_e({\bf r}_1,t)n_p({\bf r}_2,t)  
 \partial_\mp n_e({\bf r}_3,0) n_p({\bf 0},0)  
\right\rangle 
\nonumber\\
=\langle
\partial_\pm n_e({\bf r}_1,t)\partial_\mp n_e({\bf r}_3,0)
\rangle \langle n_p({\bf r}_2,t)  
 n_p({\bf 0},0)  
\rangle \, .
\label{factor}
\end{eqnarray}
Using (\ref{factor}) and taking Fourier transforms in (\ref{dipoleee}) gives
\begin{eqnarray}
\gamma_A(\omega)^\pm = {1 \over \omega \pi }\Bigr({ e^2 \over 4 \pi} \Bigr )^3 
\Bigr[{\omega \over m(\omega\pm \omega_{cp}) (\omega \mp \omega_{ce})}\Bigr ]^2 F(\omega)\, ,
\nonumber\\
\label{ft}
\end{eqnarray}
where,
\begin{eqnarray}
F(\omega)=\int d\omega_1 \,d^3 k \Bigr[{k_\bot^2  \over (k^2)^2}\Bigr ]
  \Bigr [ \Delta_e ({\bf k},\omega-\omega_1)\Bigr ] \Bigr[ \Delta _p({\bf k},\omega_1)\Bigr ]\, ,
\nonumber\\
\label{ftwhere}
\end{eqnarray}
and
\begin{eqnarray}
\Delta_{e,p} ({\bf k},\omega)=\int d^4 x \,e^{i {\bf k} \cdot {\bf x}}e^{i \omega t}
\langle n_{e,p} ({\bf x},t) n_{e,p} ({\bf 0},0)\rangle\,.
\label{delta3}
\end{eqnarray}
to be evaluated in the absence of Coulomb interactions.
We work out (\ref{ftwhere}) in two domains;  $\omega_{cp}<<T<1.5 \omega_{ce}$
and $T <<1.5 \omega_{cp}$, the first being a domain in which the electron occupancy
is mainly confined to the ground and first excited Landau
levels, with the magnetic effects on the proton being negligible. The 
second domain is one in which the electrons are strongly confined to the lowest level and the protons
are mainly confined to the ground and first excited Landau
levels. 

\subsection{Region: $\omega_{cp}<<T<1.5 \omega_{ce}$ }
In this domain the free proton correlator is simply, 
\begin{eqnarray}
\Delta_p({\bf k},\omega)=2 \pi n^{(0)}_e \delta (\omega)\, .
\label{prot}
\end{eqnarray}
There are collective effects that modify this correlator to which we return later.
We introduce  the variables, 
\begin{eqnarray}
&\xi_0= \sqrt {{m \beta \over 2}}\Bigr ({\omega \over k_\|}-{k_\| 
\over 2 m}\Bigr ) ~~~,~~~
\xi_1= \sqrt {{m \beta \over 2}}\Bigr ({\omega-\omega_{ce} \over k_\|}-{k_\| 
\over 2 m}\Bigr )\, ,
\nonumber\\
& \xi_{-1}= \sqrt {{m \beta \over 2}}\Bigr ({\omega+\omega_{ce} \over k_\|}-{k_\| 
\over 2 m}\Bigr ) \, ~~~,~~~ \zeta={k_\bot^2 \over 2 m \omega_{ce}} 
\label{variables1}
\end{eqnarray}
Following the rules given in the appendix we have calculated
the terms in the electron correlator that come from the first two Landau
levels only,
\begin{eqnarray}
&\Delta_e({k_\|,k_\bot},\omega)=(1-e^{-\beta \omega_{ce}})
\sqrt{2 \pi m \beta}\,n_e^{(0)} |k_\||^{-1}
e^{-\zeta}
\nonumber\\
&\times \Bigr [\zeta \exp (-\xi_1^2) 
+[1+(1-\zeta)^2e^{-\beta \omega_{ce}}]  \exp (-\xi_0^2)  
\nonumber\\
 &+e^{-\beta \omega_{ce}}\zeta \exp (-\xi_{-1}^2)  \Bigr ] \, .
\nonumber\\
\label{ecor}
\end{eqnarray}
Substituting (\ref{ecor}) and (\ref{prot}) into (\ref{ftwhere}), 
introducing a variable $s=k_\bot^2/k_\|^2$, and then doing the $k _\|$ integration, we 
obtain,
\begin{eqnarray}
F(\omega)={16\over 3}
\sqrt{2 m \beta }\,\pi^{3/2}[n_e^{(0)}]^2  \Lambda \, ,
\label{F}
\end{eqnarray}
where  $\Lambda$ has exactly the same meaning as
in ref. \cite{pc}, and is given by,
\begin{eqnarray}
&\Lambda={3 \over 4}(1-e^{-\beta \omega_{ce}})
e^{\beta \omega/2}\int_0^\infty ds{s \over  (1+s)^2}
\nonumber\\
&\times \Bigr [ K_0\Bigr (\beta 
\omega \sqrt{.25+ s/\omega_{ce} \beta}\Bigr )
 \Bigr ( 1+{3-2 s+s^2\over (s+1)^2}e^{-\beta \omega_{ce}} \Bigr ) 
\nonumber\\
&\,
\nonumber\\
&+(1+s)^{-1} e^{-\beta \omega_{ce}/2}\Bigr [ K_0\Bigr (\beta 
|\omega-\omega_{ce}| \sqrt{.25+s/\omega_{ce} \beta}\Bigr )
\nonumber\\
&+K_0\Bigr (\beta 
|\omega+\omega_{ce}| \sqrt{.25+ s/\omega_{ce} \beta}\Bigr )\Bigr]
\Bigr ]\,.
\label{fte}
\end{eqnarray}
This answer is identical to the results of Pavlov and Panov \cite{pp}, as corrected by Potekhin and Chabrier
in eqn.(44) of ref. \cite{pc}, when the latter is expanded in
powers of $\exp[-\beta \omega_{ce}]$, and only the zeroth and first order terms retained. 

\subsection{Region: $T < 1.5 \omega_{cp}$}  

We choose the region to extend to $1.5 \omega_{cp}$ in order to capture the resonance
behavior, while keeping the electrons strongly confined to the lowest Landau level. 
Therefore we take only the term with unity in the final factor in (\ref{ecor}) for the electronic correlator..
The contribution of the first two Landau levels to the proton correlator is given by taking
$m\rightarrow M$, and $\omega_{ce}\rightarrow \omega_{cp}$ in (\ref{ecor}) as it stands.
Doing the $\omega_1$ integral in (\ref{ft}), 
discarding terms of relative order $m/M$, and setting $\exp (-\beta \omega_{ce})=0$, appropriate to the temperature regime, we obtain,
\begin{eqnarray}
&F(\omega)=(1-e^{-\beta \omega_{cp}})
\sqrt{2 m \beta }\,\pi [n_e^{(0)}]^2  
\int  {d^3 k \over|k_\||}\Bigr[{k_\bot^2  \over (k_\|^2+k_\bot^2)^2}\Bigr ]
\nonumber\\
&\times e^{-2 \zeta}\exp \Bigr[{\beta \omega \over 2}-{k_\|^2 \beta \over 8 m }\Bigr ]
\Bigr\{\Bigr( 1+e^{-\beta \omega_{cp}}(1-\zeta)^2\Bigr )
\nonumber\\
&\times \exp\Bigr[-{\omega^2 m \beta \over 2 k_\|^2 }\Bigr]
+\zeta e^{-\beta \omega_{cp}/2}\exp\Bigr[-{(\omega-\omega_{cp})^2 m \beta \over 2 k_\|^2} \Bigr ]
\nonumber\\
&+
\zeta e^{-\beta \omega_{cp}/2 }\exp\Bigr[-{(\omega+\omega_{cp})^2 m \beta \over 2 k_\|^2} \Bigr ] 
\Bigr \}\, .
 \label{ftp}
\end{eqnarray}
In the calculation
we replaced the reduced mass by $m$ in several places.
Then the only place the proton mass enters is through $\omega_{cp}$. Note that
$\zeta=k_\bot^2 /(m \omega_{ce})=k_\bot^2 /(M \omega_{cp}) $. 
Doing the $k_\|$ integration
we obtain (\ref{F}) with $\Lambda$ replaced by $\Lambda'$, where 
\begin{eqnarray}
&\Lambda'={3 \over 4} (1-e^{-\beta \omega_{cp}})
e^{\beta \omega/2}\int_0^\infty ds{s \over (1+s)^2}
\nonumber\\
&\times \Bigr [ K_0\Bigr (\beta 
\omega \sqrt{.25+2 s/\omega_{ce} \beta}\Bigr )
\Bigr ( 1+{2 s^2+1 \over 2(s+1)^2}e^{-\beta \omega_{cp}} \Bigr ) 
 \nonumber\\
&+(1+s)^{-1}e^{-\beta \omega_{cp}/2}K_0\Bigr (\beta 
|\omega-\omega_{cp}| \sqrt{.25+2 s/\omega_{ce} \beta}\Bigr )
\nonumber\\
&+(1+s)^{-1}  e^{-\beta\omega_{cp}/2}K_0\Bigr (\beta 
|\omega+\omega_{cp}| \sqrt{.25+2 s/\omega_{ce} \beta}\Bigr )
\Bigr ] \,.
\nonumber\\
\label{ft8}
\end{eqnarray}

Note the very close resemblance to (\ref{fte}), even though (\ref{ft8})  is to be used in a domain of temperature 1000 times
smaller, at a given magnetic field. Of course when we took the two lowest Landau states for the case
of quantized protons, rather than for electrons, the energy difference $\omega_{ce}$ in (\ref{fte})
is replaced by $\omega_{cp}$ in (\ref{ft8}). But note that it is still the electronic parameter $\omega_{ce}$
that enters the $\sqrt{.25+2s/\omega_{ce}}$ factor in the arguments of the Bessel functions,
, but with a coefficient
that is different by a factor of 2. When the
temperature is so low that we have $\omega_{ce} \beta=2000$, then the integral of the
first $K_0$ function in (\ref{ft8}) becomes rather large, since convergence for large $s$ comes from
the cutoff supplied by the $K_0$ function. 

Potekhin and Chabrier \cite{pc} have given formulae which should exactly agree with (\ref{ft8}).  
However, they went quite a different route to obtain these formulae and end up with 
complex expressions that we have not been able to cast into our form. However, we obtain a 
plot similar that of the lower panel in their fig. 6. showing the peak in the scattering rate for the
anti-resonant polarization, when plotted over the proton resonance region.

\section{Electron resonance region.}
We now address aspects of behavior in the resonant region that we believe are not adequately treated
in the present literature. 
Specializing to the case of our first domain
$\omega_{cp}<<T<1.5 \omega_{ce}$ appropriate to the electron resonance region, (\ref{ft}) becomes
\begin{eqnarray}
\gamma_A(\omega) = {e^6 \over 64 \pi^4 \omega  }
\Bigr[{1\over  m (\omega-\omega_{ce})}\Bigr ]^2 F(\omega)\, ,
\label{ftwhatever}
\end{eqnarray}
an expression  that
doesn't exist at the resonance frequency; we must turn to the $\omega\rightarrow \omega-i \epsilon$  prescription
for the definition. The way that this works is that we first recognize (\ref{ftwhatever}) for the non-resonant 
case as the imaginary part of an amplitude defined by the 
graph of fig 2.
\begin {figure}[ht]
    \begin{center}
       \epsfxsize 3 in
        \begin{tabular}{rc}
           \vbox{\hbox{
$\displaystyle{ \, { } }$
               \hskip -0.1in \null} %\vskip -3 in
} &
            \epsfbox{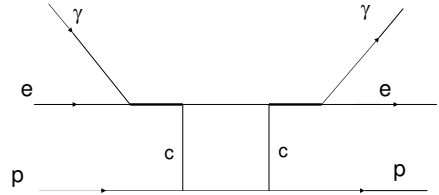} \\
            &
            \hbox{} \\
        \end{tabular}
    \end{center}
\label{fig. 2}
\vskip -.5 in
\protect\caption
    {The basic $\gamma +e \rightarrow \gamma +e$ graph, the imaginary part of which
gives the resonant part of the photoabsorption rate. The heavy lines indicate the first excited Landau level.}
\end {figure}

Accordingly, we replace 
 (\ref{ftwhatever}) by
\begin{eqnarray}
\gamma_A(\omega) = {e^6 \over 64 \pi^4 \omega  }
{\rm Im}\Bigr [ \Bigr ( {1\over  m (\omega-\omega_{ce} )}\Bigr )^2 G(\omega-i \epsilon)\Bigr]\, .
\label{ftwhatever2}
\end{eqnarray}
where $F (\omega)={\rm Im} G(\omega-i \epsilon)$.
When we move into the resonance region the denominator factor $ (\omega-\omega_{ce} )$
must have an imaginary part as well. In our original definition of the problem in which
the only photons are the external ones that coupled to the currents in 
(\ref{photoabs}) this imaginary
part originates in a self energy insertion, $\Sigma^{(ff)}$ in the inverse propagator
of the resonance. The imaginary part, ${\rm Im} \Sigma^{(ff)}$ comes from our earlier
calculation, but redescribed
(with the appropriate multiplying coefficients) as the collisional deexcitation rate, $\nu_{ff}$ for the 
resonant state. Of course, the resonance inverse propagator also has an imaginary part coming
from the intermediate state in which a photon reappears, defining the radiative (or natural) width $\nu_{re}=(2/3)e^2 \omega_{ce} \omega m^{-1}$ in the
absence of Coulomb collisions. We denote the sum of the two imaginary parts, ${\rm Im} [\Sigma^{(ff)}
+\Sigma^{(r)}]$, by
$\nu_e$, where $\nu_e=\nu_{ff}+\nu_{re}$. 

It is conceptually incorrect simply to replace $(\omega-\omega_{ce} )^{-2}$
in (\ref{ftwhatever2}) by $[(\omega-\omega_{ce} )^{2}+\nu_e^2]^{-1}$. It also can lead to
very incorrect numerical results, depending on the magnitude of 
${\rm Re}[G]$. Instead
(\ref{ftwhatever2}) should be replaced by   

\begin{eqnarray}
\gamma_A(\omega)={e^2 n_e^{(0)}  \omega_{ce}\over \omega  m}{\rm Im}[\omega-\omega_{ce}-\Sigma^{(r)}-
\Sigma^ {(ff)}]^{-1} \, ,
\label{prop1}
\end{eqnarray}
where $\nu_e={\rm Im}[\Sigma^{(r)}+\Sigma^{(ff)}]$. The prefactor in (\ref{prop1}) is
determined by expanding to first order in ${\rm Im}\Sigma^{(ff)}=\nu_{ff}$ and
comparing to (\ref{ftwhatever}), with the identification, 
\begin{eqnarray}
\nu_{ff}={e^4 {\rm Im}\Sigma^{(ff)}(\omega) \over 64 \pi^4m \omega_{ce} n^{(0)}_e}={\alpha^2 F(\omega) \over 4 \pi^2 m \omega_{ce} n^{(0)}_e} \, .
\label{prop}
\end{eqnarray} 
A graphical representation of (\ref{prop1}) is shown in fig. 2. 
\begin {figure}[ht]
    \begin{center}
       \epsfxsize 3.75in
        \begin{tabular}{rc}
           \vbox{\hbox{
$\displaystyle{ \, { } }$
               \hskip -0.1in \null} %\vskip 0.2in
} &
            \epsfbox{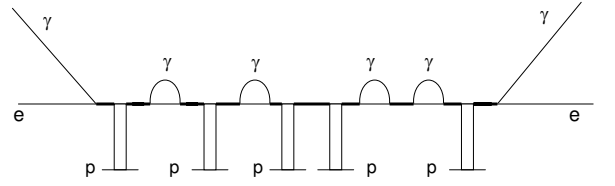} \\
            &
            \hbox{} \\
        \end{tabular}
    \end{center}
\label{fig. 3}
\vskip -1.5 in
\protect\caption
    {  A typical graph that enters the photoabsorption calculation
in the resonance region. The incoming photon is absorbed on an
electron in the lowest Landau level (exciting the electron to
the next Landau level). The initial state is restored at
the right hand side of the diagram. In between we have the propagator
for the resonance, as given in (\ref{prop}), with self energy insertions
corresponding both to the ``free-free" scattering from protons
and from free decay into photons as well. The photoabsorption rate
is found from the imaginary part of the sum of all such graphs.}
\end {figure}
 
The result (\ref{prop1}), where we use $\Sigma^{(r)}+\Sigma^{(ff)}=i(\nu_{ff}+\nu_{re})$,
gives exactly the usual results
\cite{pc} for the resonant region. Now, however, we can ask about the
effects of $\Sigma^{(ff)}$ and $\Sigma^{(r)}$
If these real parts of
 are small and not too energy dependent, then they provide
small shifts in the resonance energy that make no difference either to the total
rate or to the shape of the spectrum. But it turns out that $\Sigma^{(ff)}(\omega)$
has a logarithmic singularity in its real part at $\omega=\omega_{ce}$ that is 
of the same nature as that in its imaginary part, and we will see that it can no
longer be dismissed. 

In calculating ${\rm Re}\Sigma^{(ff)}$, we can avoid introducing a formalism with
propagators for the resonant state by directly writing
\begin{eqnarray}
\Sigma^{(ff)} ={\alpha^2  \over  \pi m \omega_{ce}} \int \,d^3 k \Bigr[{k_\bot^2  \over (k^2)^2}\Bigr ]
  \Bigr [ {[\Pi_e({\bf k},\omega)]\over (1-e^{-\beta \omega})}\Bigr ]  ,
\nonumber\\
\label{blah1}
\end{eqnarray}
where we have used (\ref{ftwhere}) and (\ref{prot}), and have extended $\Delta ({\bf k},\omega)$
of (\ref{ftwhere}) to the complex plane with,

\begin{eqnarray}
\Delta ({\bf k},\omega)= {{\rm Im}[\Pi_e({\bf k},\omega)]\over (1-e^{-\beta \omega})}\,,
\label{blah2}
\end{eqnarray} 
The imaginary part of (\ref{blah1}) reproduces the above results for $\nu_{ff}$.
The function $\Pi_e({\bf k},\omega)$ is the Fourier transform
of the retarded commutator (see ref.\cite{fw}, sec. 33), 
\begin{equation}
\Pi_{e}({\bf r},t)=\langle [n_{e}({\bf r},t) , n_{e}(0,0)]\rangle \theta(t)\,.
\label{defpi}
\end{equation}
which supplies the extension
to the complex plane that has the correct analytic properties.

To construct the real part of $\Pi$ we extend the variable set (\ref{variables1}),

\begin{eqnarray}
&\xi_0^\pm= \sqrt {{m \beta \over 2}}\Bigr ({\omega \over k_\|}\pm{k_\| 
\over 2 m}\Bigr ) ~~,~~
\xi_1^\pm= \sqrt {{m \beta \over 2}}\Bigr ({\omega-\omega_{ce} \over k_\|}\pm{k_\| 
\over 2 m}\Bigr )\, ,
\nonumber\\
& \xi_2^\pm= \sqrt {{m \beta \over 2}}\Bigr ({\omega+\omega_{ce} \over k_\|}\pm{k_\| 
\over 2 m}\Bigr )  ~~,~~ \zeta={k_\bot^2 \over 2 m \omega_{ce}}\, ,
\label{variables2}
\end{eqnarray}
and we define the usual plasma function,
\begin{eqnarray}
 \Phi(\xi )=2 e^{-\xi^2} \int_0^\xi dy \,e^{y^2} \, .
\end{eqnarray}
Then we find,
\begin{eqnarray}
&{\rm Re}[\Pi_e ({\bf k},\omega)]
={\sqrt{{m \beta}}\,n_ee^{-\zeta} (1-e^{-\beta \omega_{ce}})
\over \sqrt 2k_\|}
\nonumber\\
&\times \Bigr [ \Bigr ( \Phi(\xi_0^-)-\Phi( \xi_0^+)\Bigr )
\Bigr (   1+e^{- \beta \omega_{ce}} (1-\zeta)^2 \Bigr )
\nonumber\\
&+ \zeta
\Bigr (\Phi(\xi_1^-)- \Phi(\xi_2^+) \Bigr )
 +\zeta e^{-\beta \omega_{ce}} \Bigr (\Phi(\xi_2^-)- \Phi(\xi_1^+)\Bigr )
 \,\Bigr ]~.
\nonumber\\ \,
\label{polarization}
\end{eqnarray} 

With the real part of $\Sigma^{(ff)}$ of (\ref{blah1}) determined from
(\ref{polarization}) and the imaginary part from (\ref{blah2}) the absorption
rate is calculated from 
from (\ref{prop1}).
In fig. 3 we show the difference that inclusion of the real part
of the resonance self energy term can make in a region near the
resonance peak., for the case of parameters, $\rho=10 gc^{-3}$,
$\omega_{ec}=1$ KeV,$T=.2$KeV. Plotted are the opacities in a region
$1.01$KeV $<\omega<1.1$KeV, coming from (\ref{prop1}) using the real part
of $\Sigma^{(e)}$ that comes from (\ref{polarization}) and for the
imaginary part simply $\nu_{ff}$ as computed in section 3. Neither
the real nor the imaginary parts of $\Sigma^{(re)}$ make a significant 
contribution in this region.

\begin {figure}[ht]
    \begin{center}
       \epsfxsize 2.75in
        \begin{tabular}{rc}
           \vbox{\hbox{
$\displaystyle{ \, { } }$
               \hskip -0.1in \null} %\vskip 0.2in
} &
            \epsfbox{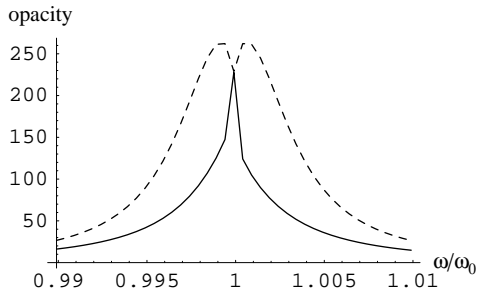} \\
            &
            \hbox{} \\
        \end{tabular}
    \end{center}
\label{fig. 4}
%\vskip 1in
\protect\caption
    { Comparison of the absorption rates, $\gamma_A$, very near the electron resonance with and
without inclusion of the real part of the self energy part in the
resonance propagator. The dashed curve is the conventional model
with no real part. The resonance energy is $.2$ KeV and the plot is
from 1\% below the resonance energy to one percent above.}
\end {figure}. 

\section{Proton resonance region}

With these results of the previous section in mind, we return to the case in which both proton and electron
currents contribute, where the interference term is important. We will consider the $^{(+)}$
polarization state, which resonates with the proton at energy $\omega_{cp}$,
but never with the electron, and stay in the energy region $\omega<<\omega_{ce}$
throughout. We go back to (\ref{ft}), rewritten to separate the proton and electron current as in
(\ref{2com}), 

\begin{eqnarray}
\gamma_A(\omega)={e^6 F(\omega) \over  \omega (4 \pi)^3} \Bigr({1 \over M(\omega- \omega_{cp})}
+{1\over m(\omega + \omega_{ce})}\Bigr )^2
\nonumber\\
\approx {e^6 F(\omega) \over  \omega M^2 (4 \pi)^3}\Bigr [\omega_{cp}^{-2}+ 2 \omega_{cp}^{-1}(\omega-\omega_{cp})^{-1} 
+(\omega-\omega_{cp})^{-2}\Bigr ]\, ,
\nonumber\\
\label{factor}
\end{eqnarray}
where we have used $m \omega_{ec}=M\omega_{pc}$, and $\omega<<\omega_{ce}$ to get the second form.

In the first term the photon is absorbed on the electron; since we are totally out
of the resonance region we do not need to build it into the propagator for
the electron resonance, as above, although there would be no harm in doing so. 
We evaluate it as is, denoting the contribution as $\gamma_{(1)}$,

\begin{eqnarray}
\gamma_{(1)}={e^6 F(\omega) \over \pi (4 \pi)^3 \omega M^2 }\omega_{cp}^{-2}\, .
\end{eqnarray} 
The cross term, that is, the interference between the amplitude in which a photon is
absorbed on the electron and that in which the photon is absorbed on the proton, is trickier. It is not a constant times the imaginary
part of a self-energy in a resonance propagator; we evaluate it almost as it stands. We recognize that it is a piece of an imaginary part of
a function of something with four legs, but save ourselves the chore of further formal definitions, by noting that
the function $F(\omega)$ defined in (\ref{ftwhere}) is the imaginary part of
an analytic function, $G(\omega)$, the real part of which we have calculated, but which is insignificant in what follows. The contribution of the second (cross) term in (\ref{factor})
becomes, 

\begin{eqnarray}
\gamma_{(2)}(\omega)= {2 e^6 \over (4 \pi)^3 \omega  \omega_{cp} M^2}{\rm Im}
 [  {G(\omega)\over \omega-\omega_{cp} -i \nu_{p} (\omega)}]
\nonumber\\
\approx  {2 e^6 \over (4 \pi)^3  \omega  \omega_{cp} M^2}   {(\omega-\omega_{cp} )F(\omega)\over (\omega-\omega_{cp})^2 +\nu_{p} (\omega)^2} \, ,
\label{ft1}
\end{eqnarray}
where $\nu$ is just the imaginary part of the proton-resonance  inverse propagator under the influence both of the magnetic
field and the Coulomb scattering, $\nu_p=\nu_{rp}+\nu_{ff}$ with  $\nu_{rp}=(2/3) e^2 M^{-1} \omega \omega_{cp}$.
In the second line of (\ref{ft1}) we have set the real part of $G$ equal to zero after all, after calculating it approximately
and finding it inconsequential. (It was important to have it included in principle, since our calculation should always
be that of calculating the imaginary part of an analytic function.)

As we found in the last section the free-free width term in the denominator $\nu_{ff}$ can be expressed in terms of
the function $F$ of (\ref{ftwhere})as 
\begin{eqnarray}
\nu_{ff}(\omega)={\alpha^2 F(\omega)\over 4 \pi^2 m \omega_{ce} n^{(0)}_e}\, ,
\end{eqnarray}

where for this case  $F(\omega)$ is evaluated as

\begin{eqnarray}
F(\omega)={16\over 3}
\sqrt{2 m \beta }\,\pi^{3/2}[n_e^{(0)}]^2  \Lambda' \, ,
\label{F2}
\end{eqnarray}
where $\Lambda '$ is evaluated in  (\ref{ft8}).

The third
term, in (\ref{factor}) leads to a contribution $\gamma_A(\omega)_{(3)}$ where the factor $(\omega-\omega_{cp})^{-2}$
needs to be fitted into the propagator for the proton resonant state, exactly
in the fashion used in the pure electron problem to obtain (\ref{prop1})

\begin{eqnarray}
\gamma_{(3)}={e^2 n_e^{(0)}  \omega_{cp}\over \omega  M}{\rm Im}[\omega-\omega_{cp}-
i \nu_{rp}-i\nu_{ff}]^{-1}\, .
\label{prop2}
\end{eqnarray}

We can now add the three contributions, replacing (\ref{factor}) by,
$\gamma_A\rightarrow \gamma_{(1)}+\gamma_{(2)}+\gamma_{(3)}$,
\begin{eqnarray}
\gamma_A(\omega) =   {e^2 n_e^{(0)}\over  \omega \omega_{cp}M} \Bigr [ {\nu_{ff} \omega^2 + \nu_{rp}  \omega_{cp}^2
\over (\omega-  \omega_{cp})^2+\nu_p^2)}
\Bigr ]\,.
\label{pcase}
\end{eqnarray}

Again we appear to disagree with the results of ref. \cite{pc}. According to eq. 53 of
that paper, the free-free part of the damping term corresponding to $\nu_p$ as it occurs
in the denominator of (\ref{pcase}), above, carries an additional factor of $m/M$
compared to our expression. Also the second term in the numerator of (\ref{pcase})
is missing in ref. \cite{pc}.

\section{Collective effects}
For problems involving Coulomb forces in a plasma, screening is
the leading collective effect. Indeed the authors
of ref. \cite{pl}, and previous authors as well, in effect replace the factor 
$[k^2]^{-2}$ in (\ref{ftwhere}) by the Fourier transform of
a screened potential, $[k^2+\kappa_D^2]^{-2}$, where $\kappa_D$
is the usual screening parameter.
This replacement is already known to be incorrect in the $B=0$ case; the correct static screening correction for photo-absorption in a hydrogen plasma
has been shown \cite{bekefi} instead to be the replacement, 
\begin{equation}
k^{-4}\rightarrow k^{-2} {1+\kappa_D^2/(2 k^2 )\over k^2+\kappa_D^2}\,.
\end{equation}
We expect the same result for the magnetic case; 
in accord with Sitenko's conclusions \cite{sitenko}, but we shall nonetheless
look at the matter in some detail, as there are non-static
corrections that may be significant in some regions.
We return to (\ref{ftwhere}) but re-express the right hand side
using (\ref{defpi}) to obtain,
\begin{eqnarray}
&F(\omega)=\int d\omega_1 \,d^3 k \Bigr[{k_\bot^2  \over k^4}\Bigr ] {{\rm Im} \Bigr [\Pi_e' ({\bf k},\omega-\omega_1)\Bigr ]\over
 (1-e^{-\beta (\omega-\omega_1)})}{ {\rm Im}\Bigr[ \Pi_p'({\bf k},\omega_1)\Bigr ]\over 
(1-e^{-\beta \omega_1})} \, ,
\nonumber\\
&\,
\label{whatever2}
\end{eqnarray}
where the $\Pi'$ are the functions defined by (\ref{defpi}), but now
calculated, in approximation, in the presence of Coulomb forces, whereas the
the functions, $\Pi$ are the functions defined in (\ref{defpi})in the absence
of Coulomb forces. In the ring approximation, we have,
\begin{eqnarray}
\Pi'_e({\bf k},\omega)={[k^2+4 \pi e^2 \Pi_p ({\bf k},\omega)]\Pi_e({\bf k},\omega) \over
k^2+ 4 \pi e^2 \Pi_e({\bf k,\omega})+4 \pi e^2\Pi_p(\bf k,\omega)} \,,
\nonumber\\
\Pi'_p({\bf k},\omega)={[k^2+4 \pi e^2\Pi_e ({\bf k},\omega)]\Pi_p({\bf k},\omega) \over
k^2+ 4 \pi e^2 \Pi_p({\bf k,\omega})+4 \pi e^2 \Pi_e(\bf k,\omega)}\,.
\label{ring}
\end{eqnarray}
We can we describe this construction as the expression of
the complete polarization parts $\Pi_e', \Pi_p'$
from the proper polarization parts, $\Pi_e, \Pi_p$, then setting the proper parts to  
their values in the absence of Coulomb interactions.
The best systematic derivation that we know of for these relations 
in a multicomponent classical plasma is in ref. \cite{brown and yaffe},
eq. 2.110. 

The relations hold in the presence of quantum effects as well.
We have obtained tractable
expressions from substituting the results of (\ref{ring})
into (\ref{whatever2}) only for the first case of section 3, 
$\omega_{cp}<<T<1.5 \omega_{ce}$, where we can neglect the magnetic
interactions of the protons.
For this case
the large proton mass 
leads to the imaginary part of the proton 
correlator ${\rm Im}[D_p]$ in (\ref{ft}) being concentrated at very small values
of $\omega_1<<T$, for relevant values of $k\approx \sqrt {mT}$. Thus in (\ref{ft}) we can set
$\omega-\omega_1=\omega$, and $(1-\exp[-\beta \omega_1])^{-1}=(\beta \omega_1)^{-1}$,
and use the dispersion relation to do the $\omega_1$ integral,
\begin{eqnarray}
&\pi ^{-1}\beta^{-1} \int d\omega_1 {\rm Im} \Pi_p' ({\bf k,\omega_1})/\omega_1= 
\beta^{-1}{\rm Re}\Pi_p' ({\bf k},0)
\nonumber\\
&={4 \pi e^2 n_e^{(0)} (1+\kappa^2_e/k^2)\over 
(1+\kappa^2_p/k^2+\kappa_e^2/k^2)}\, ,
\label{ftx}
\end{eqnarray}
where we have simplified by setting
\begin{eqnarray}
4 \pi e^2 \Pi_{e,p}({\bf k},0)\approx 4 \pi e^2 \Pi_{e,p}(0,0)=\kappa^2_{e,p}\, .
\end{eqnarray}
Here the $\kappa^2_{e,p}$ are the contributions of the individual species to the
squared Debye wave number $\kappa_D^2=\kappa^2_e+\kappa^2_p$ 
and $\kappa^2_p=\kappa^2_e=4\pi \beta  e^2 n^{(0)}_e $.
We have also rewritten the multiplying
factor of $\kappa_p^2$ in (\ref{ft1}) terms of the average proton density, $n_p^{(0)}=n_e^{(0)}$ and the
temperature, $\beta^{-1}$. For the static response function ($\omega=0$) this is a good approximation
for all cases under consideration.

Putting these steps into (\ref{ft}), using the ring approximation (\ref{ring}) for the electron polarization,
and noting that $\Pi_p(k,\omega  \approx \beta^{-1})\approx 0$  gives 

\begin{eqnarray}
&F(\omega)={n_e^{(0)}\over  1-e^{-\beta \omega}}\int  d^3 k \Bigr[{k_\bot^2  \over k^4}\Bigr ] 
{{\rm Im} \Bigr [\Pi_e^{(0)} ({\bf k},\omega)\Bigr ]\over |1+\Pi_e^{(0)} ({\bf k},\omega)/k^2|^2} 
\nonumber\\
&\times{(1+\kappa^2_e/k^2)\over 
(1+\kappa^2_p/k^2+\kappa_e^2/k^2)}\, .
\label{ftx}
\end{eqnarray}

We can compare (\ref{ftx}) for the case $B=0$ with results in the literature giving the effects of Coulomb correlations
on the photo-absorption rate. Taking the appropriate limits of the prefactor in (\ref{dipoleee}), substituting the
$B=0$ form for  ${\rm Im }\Pi_e^{(0)}/k^2$, and defining $\epsilon (k,\omega) =1+
\Pi_e^{(0)}(k,\omega)/k^2$, and including the longitudinally polarized modes gives back exactly eq. (3) in the paper by Iglesias and Rose \cite{ir},
and essentially the results of ref. \cite{tsytovich} as well.\footnote{ There is one  discrepancy; in both
of these references the dielectric function $\epsilon$, where it occurs in the denominator, is taken to be that of a classical plasma. The numerators,
which are not in these works identified as the imaginary part of the dielectric functions, require the quantum
treatment in order to avoid an ultraviolet divergence (or the introduction on an arbitrary logarithm). In any application
the full quantum form should probably be used in the denominator as well.} The corrections from the
$|\epsilon|^2$ in the denominator and from the ionic correlator are actually relatively small in domains of density and temperature
in which the plasma is weakly coupled, that is to say, in regions in which we can calculate at all. We note that the last factor on the right hand side of
(\ref{ftx}) provides the screening factor that we quoted at the beginning
of this section.

\section{Discussion}
We summarize our differences from previous authors in three different parameter 
regions:
\subparagraph{1. Region of non-quantized protons.}
Here we recapture the results of other authors for the ``logarithm",
or ``Gaunt factor", $\Lambda$.  But there is another effect, at the same 
level of approximation, that can change the shape of the resonant peak
by a lot, namely the rapid energy dependence of the real part
of the self energy in the propagator for the resonance.
Because of this energy dependence the real part is not merely
a small adjustment to the resonant energy parameter, and can make
large modifications to both the spectrum shape and the integral
over the resonance region. We believe that the effect will be important
in the proton-resonance region as well, but we have not calculated examples.

\subparagraph{2. Region very close to the proton resonance.}
While we agree with the authors of ref. \cite{pc} in the way that the free resonance decay partial width parameter $\nu_{rp}$
enters the total width parameter $\nu_p$, we appear to disagree with these authors on the
contribution of the free-free parameter $\nu_{ff}$ itself to this width.
\subparagraph{3. Collective effects} We have included the basic ionic screening effects
which are somewhat different from those of other authors, and which, in contrast, 
reduce to well known collective corrections to photoabsorption in the limit of no magnetic field.

\section{Appendix}
The electron field operator $\psi({\bf r},t)$ is built in cylindrical coordinates from the states of the theory, in
the usual way,
\begin{eqnarray}
&\psi({\bf r},t)=\sum_{n,s}\sum_p L^{-1/2} a_{n,s}(p) e^{ipz }
\nonumber\\
&\times e^{-i(p^2/2m+n \omega_c )t} 
 u_{n,s}({|\bf r}_\bot|) e^{-i(n- s)\phi}
\label{fields}
\end{eqnarray}
where $p$ is the momentum in the direction of the field $ a_{n,s}(p)$ is the annihilation operator for the indicated mode, and $\epsilon_p=p^2/2m$. To calculate
the correlator needed in (\ref{delta3}),

\begin{equation}
\Delta_e({\bf x},t)=\langle \psi^\dagger({\bf r},t)\psi({\bf r},t)
 \psi^\dagger({\bf 0},0)\psi({\bf 0},0)\rangle \,,
\label{cor6}
\end{equation}
we first write
\begin{eqnarray}
\psi ({\bf r},t)\psi^\dagger ({\bf 0},0)=-\psi^\dagger ({\bf 0},0)\psi ({\bf r},t)+C({\bf r},t)
\label{com}
\end{eqnarray} 
where the anticommutator function $C$ is a c-number. In the present work we are considering only non-degenerate
electrons and therefore we discard the first term on the RHS of (\ref{com}), since its contribution is of higher order in the fugacity 
of the electrons, $e^{\beta \mu_e}$, i.e., a correction for Fermi statistics. (Note that the function $C$ itself would
be unchanged for the case of Bose statistics, while the discarded term would be of the other sign.)

Before proceeding
further we note that since the functions $u_{n,s}({\bf r}_\bot)$ for $n\ne s$ vanish at ${\bf r}_\bot)$, and one electron field is
evaluated at ${\bf r}_\bot)$ in each expression that we encounter, only terms with $n=s$ will enter, and the azimuthal angle
$\phi$ will not appear in any expressions. Henceforth we use radial functions labeled with the index $n$ alone;
$u_n(\rho)\equiv u_{n,n}(\rho)$.

Explicitly, the function $C$ is now given by

\begin{eqnarray}
C(|{\bf r}_\bot| ,z,t)= \sum_n \int{dp \over 2 \pi}e^{ipz}e^{-i [p^2/(2 m)+n \omega_c]t}u_n(|{\bf r}_\bot |)u_n(0)
\nonumber\\
\,
\label{C}
\end{eqnarray}
For the remainder of the evaluation of (\ref{cor6}) and its Fourier transform we need the thermal expection value,
\begin{eqnarray}
\langle [a_n(p)]^\dagger [a_n(p')]\rangle=\delta_{n,n'}
\delta_{p,p'}e^{\beta \mu_e}
e^{-(p^2/2m+n \omega_c)\beta}\, ,
\label{thermal}
\end{eqnarray}

where to determine the fugacity $\exp[\beta \mu_e]$ to be used in (\ref{thermal}) we calculate the electron
density (for convenience at $r=0$) as
\begin{eqnarray}
n_e=e^{\mu_e \beta}\sum_{n=0}^\infty |u_n(0)|^2 e^{-n \beta \omega_c}\int_{-\infty}^\infty {d p \over 2 \pi}
e^{-\beta p^2 /(2m)}
\end{eqnarray}
Using $|u_n(0)|^2=m\omega_c/(2\pi)$ we obtain,
\begin{eqnarray}
e^{\beta \mu_e}=(1-e^{-\beta \omega_{ce}}){\pi^{3/2} 2^{3/2}n_e^{(0)}\beta^{1/2} \over m^{3/2}\omega_{ce}}\,
\label{fug}
\end{eqnarray}
Putting together (\ref{com}), (\ref{C}), and (\ref{delta4})
we obtain
\begin{eqnarray}
\Delta_{e} ({\bf k},\omega)=e^{\beta \mu_e}\int d^2r_\bot \, dz \,e^{i {\bf k_\bot} \cdot {\bf r_\bot}}e^{i \omega t}e^{ik z}
\nonumber\\
\times \sum_{j= 0}^\infty \int {dp \over 2\pi}e^{i p z} u_j(|{\bf r_\bot}|)u_j(0)C({\bf r}_\bot,z,t)
\label{delta4}
\end{eqnarray}
For examination of behavior in the region of the main cyclotron resonance it suffices to include only the terms
$j=0,1$ in the sum in (\ref{delta4}) and terms with $n=0,1$ in the sum in(\ref{C}). Inserting 
\begin{eqnarray}
u_0(\rho)=\sqrt{{\omega_c m \over 2 \pi}}e^{-m \omega_c \rho^2/4}~,
\nonumber\\
 u_1(\rho)={m\omega_c\over 2{\sqrt \pi}}\rho e^{-m \omega_c \rho^2/4}
\end{eqnarray}
where $\gamma=\omega_c m/2$,
and performing the Fourier transforms yields the result (\ref{ecor}).


\begin{thebibliography}{99}
\bibitem{pp} G.G. Pavlov and A. N. Panov, JETP 71,572 (1976)[Sov.Phys. 44,300]
\bibitem{mesz} P. Meszaros, ``High Energy Radiation from Magnetized Stars"
(Chicago:University of Chicago Press, 1992)
\bibitem{vj}Virtamo, J.\& Jauho, P.  NCimB. 26,  537 (1975)	
\bibitem{nv}Nagel, W.\& Ventura, J.A \&A 118, 66 (1983)
\bibitem{pc}A. Y. Potekhin , G. Chabrier , Astrophys.J. 585, 955(2003) 
\bibitem{pp2}G. C. Pavlov,A. Y. Potekhin, Astrophys. J. 450, 883 (1995)
\bibitem{pl}A. Y. Potekhin  \& D. Lai,  Mon. Not. R. Astron. Soc. 376, 793 (2007)

\bibitem{hl1}W.C.G. Ho, \& D. Lai, MNRAS,327,1081 (2001)
\bibitem{ozel} F. Ozel, Astrophys. J. 563, 276 (2001)
\bibitem{zane1}S Zane, R. Turrola, \&A Trevis, Astrophys. J. 537, 387 (2000)
\bibitem{zane2}S Zane, R. Turrola, L. Stella \&A Trevis, Astrophys. J. 560, 384 (2001)
\bibitem{fw} A. Fetter and D. Walecka, ``Quantum Theory of Many-Particle Systems",
(New York:McGraw-Hill, 1971)
\bibitem{brown and yaffe} Brown, L. S. and Yaffe, L. G. 2001, Physics Rept. 340, 1
\bibitem{bekefi} Bekefi, G. 1966, Radiation Processes in Plasmas (New York, Wiley)
\bibitem{sitenko} A. G. Sitenko, ``Electromagnetic Fluctuations in Plasma"
(New York, Academic Press, 1967) 
\bibitem{ir}Iglesias,C. A. and Rose, S.J. 1996, ApJ,466,L115
\bibitem{tsytovich} Tsytovich, V.N.; Bingham, R.; de Angelis, U.; Forlani, A. 1996, J. Plasma Phys.  56, 127
\end{thebibliography}
\end{document}